\documentclass[12pt, prd, showpacs]{revtex4}
\usepackage{amssymb}
\usepackage{amsmath}
\usepackage{graphicx}

\begin{document}

\title{General radially moving references frames in the black hole background%
}
\author{A. \ V. Toporensky}
\affiliation{Sternberg Astronomical Institute, Lomonosov Moscow State University }
\affiliation{Kazan Federal University, Kremlevskaya 18, Kazan 420008, Russia}
\email{atopor@rambler.ru}
\author{O. B. Zaslavskii}
\affiliation{Department of Physics and Technology, Kharkov V.N. Karazin National
University, 4 Svoboda Square, Kharkov 61022, Ukraine}
\email{zaslav@ukr.net}

\begin{abstract}
We consider general radially moving frames realized in the background of
nonextremal black holes having causal structure similar to that of the
Schwarzschild metric. In doing so, we generalize the Lema\^{\i}tre approach,
constructing free-falling frames which are built from the reference
particles with an arbitrary specific energy $e_{0}$ including $e_{0}<0$ and
a special case $e_{0}=0$. The general formula of 3-velocity of a freely
falling particle with the specific energy $e$ with respect to a frame with $%
e_{0}$ is presented. We point out the relation between the properties of
considered frames near a horizon and the Banados-Silk-West effect of an
indefinite growth of energy of particle collisions. Using our radially
moving frames, we consider also nonradial motion of test particles including
the regions near the horizon and singularity. We also point out the
properties of the Lema\^{\i}tre time at horizons depending on the frame and
sign of particle energy.
\end{abstract}

\maketitle

\section{Introduction}

The Schwarzschild metric \cite{s} is a part of primer of black hole physics
and enters all textbooks on gravitation. In spite of this, it still remains
a testing area of different approaches, including classes of coordinate
transformations. As is known, this metric is singular in the original
(so-called curvature, or Schwazschild) coordinates. There exists completely
different methods to remove such a seeming singularity. These approaches can
be united in a one picture \cite{finch}. What is especially interesting is
that if the specific energy $e_{0}$ of reference particles (i.e. particle
realizing a frame) is included explicitly in the coordinate transformation,
the different standard forms can be obtained as different limiting
transitions. In this approach, one can recover some well-known metric like
the Eddington-Finkelstein ones \cite{mart}, \cite{jose}. The Lema\^{\i}tre
metric can be also included in this scheme \cite{bron}, \cite{we}.
(Alternatively, one can use a velocity of the local Lorentz transformation
instead of $e_{0}$ \cite{fom}.) Meanwhile, there exists one more aspect
connected not only with the frames themselves but with particle dynamics in
corresponding background. One may ask, how particle motion looks like
depending on $e_{0}$ and particle specific energy $e$ and relation between
them.

In the previous paper \cite{fl} we considered such frames that all reference
particles have $e_{0}=1$ or $e_{0}=0$. In the present work, we make the next
step and consider a more general situation when $e_{0}$ is arbitrary. In
particular, this includes the case of $e_{0}<0$. Motivation for such
generalization is at least three-fold. (i) In the aforementioned papers, the
introduction of $e_{0}$ was made for \ frames, now we consider particle
dynamics. (ii) If we make $e_{0}$ a free parameter, we can trace the
relation between reference particles and any other test particles thus
establishing connection between the choice of a frame and properties of
particle collisions. This is especially actual in the context of high energy
particle collisions \cite{ban}. (iii) We hope that a general approach
developed in our work will be useful tool for description of particle motion
under the horizon, some concrete examples of which in the Schwarzschild
background were discussed in \cite{and} (see also references therein).

In this work, we develop general formalism. In doing so, we suggest simple
classification of frames based on the their character (contracting or
expanding) and the sign of $e_{0}$. The applications of our formalism will
be considered in a next paper. Also, we restrict ourselves by static black
holes and postpone the generalization to rotating \ black holes to future
works.

In what follows we deal with the spherically symmetric metrics of the form%
\begin{equation}
ds^{2}=-fdt^{2}+\frac{dr^{2}}{f}+r^{2}(d\theta ^{2}+d\phi ^{2}\sin
^{2}\theta )\text{.}  \label{sch}
\end{equation}%
For the Schwarzschild metric, $f=1-\frac{r_{+}}{r}$ where $r_{+}$ is the
radius of the event horizon.

\section{Reference particles and frames}

One of known frames that removes the coordinate singularity on the horizon
is the Gullstrand-Painlev\'{e} (GP) one \cite{gul}, \cite{p}. In recent
years, this frame again attracts attention in different contexts (see \cite%
{we} and references therein). Quite recently, some modification of the GP
frame was suggested in \cite{gr}. The generalization of the original GP
frame can be obtained after introducing a new time variable via 
\begin{equation}
d\tilde{t}=e_{0}dt+\frac{dr}{f}P_{0},  \label{tt}
\end{equation}%
where by definition 
\begin{equation}
P_{0}\equiv \sqrt{e_{0}^{2}-f}.  \label{P0}
\end{equation}

The static time $t$ is expressed now as 
\begin{equation}
dt=\frac{1}{e_{0}}\left( d\tilde{t}-\frac{dr}{f}P_{0}\right)
\end{equation}%
that gives us the metric in the form 
\begin{equation}
ds^{2}=-\frac{f}{e_{0}^{2}}d\tilde{t}^{2}+\frac{2d\tilde{t}dr}{e_{0}^{2}}%
P_{0}+\frac{dr^{2}}{e_{0}^{2}}+r^{2}d\omega ^{2}.  \label{met}
\end{equation}

After introducing a new spatial variable $\rho $ via 
\begin{equation}
d\rho =\frac{dr}{P_{0}}+d\tilde{t}
\end{equation}%
the metric can be set to a synchronous form 
\begin{equation}
ds^{2}=-d\tilde{t}^{2}+\frac{P_{0}^{2}}{e_{0}^{2}}d\rho ^{2}+r^{2}d\omega
^{2}.  \label{syn}
\end{equation}

In this form it is evident that the coordinate system is formed by
particles, free falling with the specific energy $e_{0}$. Indeed, for a
radial fall with the the specific energy $e$ we have the following equations
of motion: 
\begin{equation}
\frac{dt}{d\tau }=\frac{e}{f},  \label{dtf}
\end{equation}%
\begin{equation}
\frac{dr}{d\tau }=-\sqrt{e^{2}-f}\equiv -P,  \label{dr}
\end{equation}%
where $\tau $ is the proper time.

Now we can consider the derivative 
\begin{equation}
\frac{d\tilde{t}}{d\tau }=e_{0}\frac{dt}{d\tau }+\frac{dr}{d\tau }\frac{P_{0}%
}{f}=\frac{e_{0}e-P_{0}P}{f},  \label{dt}
\end{equation}%
where $\tilde{t}$ is the $e_{0}$-synchronous time, and $\tau $ is the proper
time of a particle with the energy $e$. This entails 
\begin{equation}
\frac{dr}{d\tilde{t}}=-\frac{Pf}{e_{0}e-P_{0}P}  \label{rtt}
\end{equation}%
and 
\begin{equation}
\frac{d\rho }{d\tau }=\frac{e_{0}(eP_{0}-e_{0}P)}{f}
\end{equation}%
that gives us the rate with which the synchronous spatial coordinate $\rho $
changes for a free falling particle with the energy $e$. If $e=e_{0}$, it
follows that $P=P_{0}$ and%
\begin{equation}
\frac{d\rho }{d\tau }=0,
\end{equation}%
as is should be. We see that indeed the particle with the energy $e_{0}$ has
synchronous spatial coordinate $\rho $ constant during a free fall.

We can note also that for $e_{0}=1$ the coordinate flow velocity $dr/d\tilde{%
t}=-\sqrt{1-f}$ coincides with the velocity of a particle with respect to a
stationary frame. For a general case the velocity of the particle having $e$
with respect to a stationary frame is equal to $v_{st}=\sqrt{(e^{2}-f)/e^{2}}
$(see eqs.4.2, 4.4 and 6.5 in \cite{zero}), so that $v_{st}=P/e=(dr/d\tilde{t%
})/e$.

Let $e_{0}\rightarrow 0$. We can easily see that there is no smooth limit
for the generalized GP metric (\ref{met}). However, as it has been pointed
out in \cite{we}, we can write a regular form of the synchronous metric if
we make a redefinition $\rho =e_{0}\tilde{\rho}$. As now 
\begin{equation}
e_{0}d\tilde{\rho}=\frac{dr}{P_{0}}+d\tilde{t},  \label{ert}
\end{equation}%
we see than in this limit $r$ is the function of $\tilde{t}$ only and we get
from (\ref{syn}) a \textit{homogeneous} metric under the horizon where we
set $f\equiv -g$: 
\begin{equation}
ds^{2}=-d\tilde{t}^{2}+g(r(\tilde{t}))d\tilde{\rho}^{2}+r^{2}(\tilde{t}%
)d\omega ^{2}\text{.}  \label{g}
\end{equation}

So far, we have considered only positive energy ingoing particles and
frames. To include negative $e$ and outgoing motion, we generalize the
transformation to a synchronous frame. One can write

\begin{equation}
d\tilde{t}=e_{0}dt-\frac{dr}{f}\sigma _{0}P_{0},  \label{wave}
\end{equation}%
\begin{equation}
d\rho =\frac{dr}{P_{0}}+\delta d\tilde{t}=\frac{dr}{fP_{0}}[f(1+\delta
\sigma _{0})-e_{0}^{2}\sigma _{0}\delta ]+\delta e_{0}dt\text{.}
\end{equation}%
Here $\sigma _{0}=\pm 1,$ $\delta =\pm 1$ and we assumed that$\left( \frac{%
\partial \rho }{\partial r}\right) _{\tilde{t}}>0$.

We can introduce the function $\chi $ according to%
\begin{equation}
\chi (r)\equiv \int \frac{dr}{P_{0}}.  \label{hi}
\end{equation}%
Then, for $\delta =+1$ 
\begin{equation}
\chi =\rho -\tilde{t}  \label{hi1}
\end{equation}

and for $\delta =-1$, 
\begin{equation}
\chi =\rho +\tilde{t}\text{.}
\end{equation}

Thus for the contracting and expanding frame the dependences $r(\rho ,\tilde{%
t})$ are different.

The inverse formulas read 
\begin{equation}
dt=\frac{1}{e_{0}}(d\tilde{t}+\frac{\sigma _{0}P_{0}}{f}dr)\text{,}
\end{equation}%
\begin{equation}
dr=P_{0}d\rho -\delta P_{0}d\tilde{t}\text{.}
\end{equation}%
We want the metric in new coordinates to be diagonal, so the cross-terms
with $d\tilde{t}d\rho $ should cancel. This leads to the condition%
\begin{equation}
\delta =-\sigma _{0}.
\end{equation}%
Thus we have%
\begin{equation}
d\rho =\frac{dre_{0}^{2}}{fP_{0}}-\sigma _{0}e_{0}dt.  \label{ro}
\end{equation}

For the contracting systems, $\frac{dr}{d\tilde{t}}<0$ along the line with $%
\rho =const$, so $\delta =+1=-\sigma _{0}.$ For the expanding one, $\delta
=-1$, $\sigma _{0}=+1$. The synchronous form reads%
\begin{equation}
ds^{2}=-d\tilde{t}^{2}+\frac{P_{0}^{2}}{e_{0}^{2}}d\rho ^{2}+r^{2}d\omega
^{2}  \label{sync}
\end{equation}%
both for the contracting and expanding frames.

For a general radial motion of a geodesic particle with a specific energy $e$
the equations of motion give us (\ref{dtf}) and%
\begin{equation}
\frac{dr}{d\tau }=\sigma P\text{,}  \label{rs}
\end{equation}%
where $\sigma =\pm 1$ depending on the direction of motion.

Correspondingly, we also have from (\ref{wave}) and (\ref{ro})%
\begin{equation}
\frac{d\tilde{t}}{d\tau }=\frac{ee_{0}-\sigma _{0}\sigma PP_{0}}{f}\text{,}
\label{til}
\end{equation}
\begin{equation}
\frac{dr}{d\tilde{t}}=\frac{Pf}{ee_{0}\sigma -\sigma _{0}PP_{0}}=\frac{%
P\sigma f}{Y}\text{,}  \label{rtw}
\end{equation}%
where $Y=ee_{0}-\sigma _{0}\sigma PP_{0}$,%
\begin{equation}
\frac{d\rho }{d\tau }=\frac{\sigma P}{fP_{0}}e_{0}^{2}-\frac{\sigma
_{0}e_{0}e}{f}\text{.}
\end{equation}

If $P=P_{0}$ and $e=e_{0}$,%
\begin{equation}
\frac{d\rho }{d\tau }=\frac{e_{0}^{2}}{f}(\sigma -\sigma _{0})\text{,}
\end{equation}%
\begin{equation}
\frac{d\tilde{t}}{d\tau }=\frac{e_{0}^{2}-\sigma _{0}\sigma P_{0}^{2}}{f}%
\text{.}
\end{equation}

For reference particles, by definition, we require $\frac{d\rho }{d\tau }=0$%
, $\frac{d\tilde{t}}{d\tau }=1$. This is achieved if $\sigma =\sigma _{0}$.
It is worth noting that $\sigma _{0}$ and $\sigma $ have different meanings.
The first one is the parameter of the coordinate transformation (\ref{wave}%
), the second one is the characteristic of particle motion.

\section{Classification of frames}

Since the energy enters in (\ref{sync}) explicitly only as square, this
formula includes, in fact, four different cases depending on signs of the
first and second terms in the right hand side of (\ref{wave}). For example,
the standard Lema\^{\i}tre system, being a contracting system with $e_{0}=1,$
has three counterparts: an expanding system with $e_{0}=1$, a contracting
system with $e_{0}=-1$ and an expanding system with $e_{0}=-1$. All four are
represented by the same equation (\ref{sync}), with the differences in
particular expressions for $r$ through the coordinates $\rho $ and $\tilde{t}
$.

This poses the question about the meaning of such Lema\^{\i}tre frames,
their GP counterparts and their classification. For example, the condition
for the existence of the metric (\ref{sync}) is $e_{0}^{2}>f$. In
particular, for the Schwarzschild space-time it is defined everywhere if $%
|e_{0}|\geq 1$ independently of the sign of $e_{0}$. On the other hand, in
the Schwarzschild black hole particles can have negative energy only inside
a horizon and should be included in a general scheme. (Under the horizon,
the energy and momentum mutually interchange their meaning. However, for
shortness, we call $e$ and $e_{0}$ energies in all regions of a space-time.)

\begin{figure}[tbp]
\centering
\includegraphics{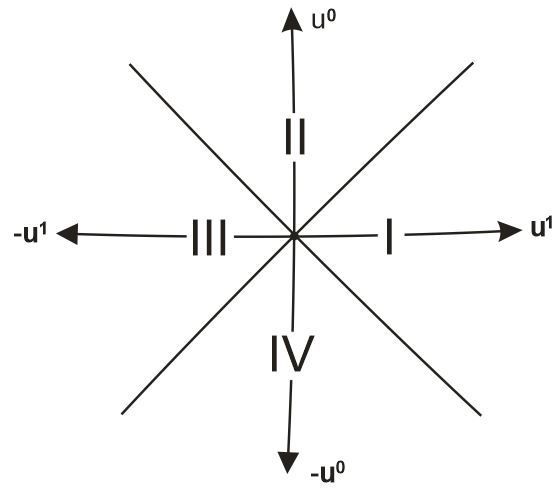}
\caption{The Kruskal diagram for the Schwarzschild metric}
\end{figure}

A possible answer is that the metric (\ref{sync}) for a negative $e_{0}$
does exist for all $r$, but in another parts of the Carter-Penrose diagram
for a geodesically complete space-time (see Fig.1). In what follows, we
consider the Schwarzschild space-time or any other space-time with a similar
structure. In doing so, we use standard notations for different regions of
space-time (see Sec. 31 in \cite{mtw}): I is "our" region, II - a black hole
region, III - a mirror one, IV is a white hole region. Then, it is
convenient to characterize a frame by two parameters. Below, $C$ means
"contracting", $E$ means "expanding. We also indicate the sign of $e_{0}$.
Then, we have 4 different frames that are listed in Table I.

\begin{tabular}{|l|l|}
\hline
Frame & Regions covered \\ \hline
$(C,+)$ & I, II \\ \hline
$(C,-)$ & III, II \\ \hline
$(E,+)$ & IV, I \\ \hline
$(E,-)$ & IV, III \\ \hline
\end{tabular}

Identification of regions follows from several facts. (i) Only particles
with $e_{0}<0$ can cross the branch of the horizon between regions III and
II. (ii) Particles that cross the branch between regions IV and I enter
"our" usual region I ($R^{+}$, according to the Novikov classification \cite%
{nov}), so they must have $e_{0}>0.$ In a similar way, particle in the
asymptotically flat mirror region III $\ (R^{-})$ must have $e_{0}<0$.
Somewhat different classification of frames was suggested in \cite{gr}.

It is worth paying attention to an important detail not noted in the
textbook \cite{LL}. The first equation in eq. 102.1 there with the lower
signs implies that reference particles have negative energy. This frame is
expanding and is valid in regions IV, III only.

Our classification applies to a full space-time of an eternal black-white
hole. However, it retains its sense in a realistic black hole also since a
particle can perform an outward geodesic motion in the $R$ region of a
realistic black hole while a particle with a negative $e$ can exist in the $%
T $ region. In principle, we can avoid to invoke the Carter-Penrose diagram,
and use the forward-in-time condition instead. Indeed, we have from (\ref%
{rtw}) that along the trajectory%
\begin{equation}
\tilde{t}=\int_{r_{0}}^{r}\frac{drY}{Pf\sigma }\text{, }
\end{equation}%
where $r_{0}$ is an initial position.

Let us consider ingoing particles$,r_{0}>r$, $\sigma =-1$. Then, 
\begin{equation}
\tilde{t}=\int_{r}^{r_{0}}\frac{drY}{Pf}\text{.}
\end{equation}

The forward-in-time condition requires 
\begin{equation}
\frac{d\tilde{t}}{d\tau }>0,  \label{ftw}
\end{equation}%
so outside the horizon ($f>0$) 
\begin{equation}
Y>0  \label{out}
\end{equation}

and inside the horizon ($f<0$) 
\begin{equation}
Y<0.  \label{in}
\end{equation}

Thus the set of admissible configurations can be described as follows.

Outside the horizon. In the $R^{+}$ region $e>0$, $e_{0}>0$, $\sigma $ and $%
\sigma _{0}$ can have arbitrary signs. In the $R^{-}$ region, $e<0$, $%
e_{0}<0 $, $\sigma $ and $\sigma _{0}$ can have arbitrary signs.

Inside the horizon. In the $T^{-}$ region $\sigma _{0}=+1$, $\sigma =-1$, $e$
and $e_{0}$ can have arbitrary signs. In the $T^{+}$ region $\sigma _{0}=-1$%
, $\sigma =+1$, $e$ and $e_{0}$ can have arbitrary signs. (Depending on
their signs, particle move from region IV to I or III.)

In what follows we omit $\sigma _{0}$ and $\sigma $ for brevity, assuming
that $P_{0}<0$ for expanding frames ($\sigma _{0}=+1$) and $P_{0}>0$ for
contracting frames ($\sigma _{0}=-1$). In a similar way, we assume that $P<0$
for outgoing particles $(\sigma =+1$) and $P>0$ for ingoing particles $%
(\sigma =-1)$.

Then, in coordinates $(\tilde{t},r,\theta ,\phi $) the four-velocity of a
free radially falling (ingoing) particle reads%
\begin{equation}
u^{\mu }=(\frac{e_{0}e-P_{0}P}{f},-P,0,0),  \label{u}
\end{equation}%
\begin{equation}
u_{\mu }=(-\frac{e}{e_{0}},\frac{P_{0}e-Pe_{0}}{fe_{0}},0,0).  \label{u2}
\end{equation}

\section{Radial 3-velocity with respect to a free falling frame}

It is well known that the 3-velocity with respect to a particular frame
characterized by a tetrad field $h_{(i)\mu }$ is given by (see, e.g. \cite%
{72})%
\begin{equation}
V^{(i)}=-\frac{h_{(i)\mu }u^{\mu }}{h_{(0)\mu }u^{\mu }}.  \label{V}
\end{equation}

We choose a radially moving particle with $e=e_{0}$ as a reference particle.
Then, $h_{(0)}^{\mu }=U^{\mu }$, where $U^{\mu }$ is its four-velocity. In
coordinates $(\tilde{t},r,0,0)$ it follows from (\ref{met}) and (\ref{u})
that%
\begin{equation}
h_{(0)}^{\mu }=U^{\mu }=(1\text{, }-P_{0},0,0),\text{ }  \label{h0}
\end{equation}%
\begin{equation}
h_{(0)\mu }=U_{\mu }=(-1,\text{ }0,0,0),  \label{hoc}
\end{equation}%
\begin{equation}
h_{(1)}^{\mu }=e_{0}(0\text{,}1,0,0),  \label{h1}
\end{equation}%
\begin{equation}
h_{(1)\mu }=\frac{1}{e_{0}}(P_{0},1,0,0).  \label{h1c}
\end{equation}%
It is worth noting that eq. (\ref{V}) can be also rewritten in the form%
\begin{equation}
V^{(i)}=\frac{h_{(i)\mu }u^{\mu }}{\gamma },  \label{Vg}
\end{equation}%
where the Lorentz factor of relative motion between a particle and an
observer $\gamma =-u^{\mu }U_{\mu }$. This expression for $\gamma $ in the
Schwarzschild background was analyzed in \cite{vm}.

Using this we can write 
\begin{equation}
-h_{(0)\mu }u^{\mu }=u^{\tilde{t}}=\gamma ,  \label{ut}
\end{equation}%
\begin{equation}
h_{(1)\mu }u^{\mu }=\frac{P_{0}u^{\tilde{t}}+u^{r}}{e_{0}}  \label{ur}
\end{equation}%
which results in 
\begin{equation}
V^{(1)}=\frac{\left( P_{0}+\frac{dr}{d\tilde{t}}\right) }{e_{0}}\equiv v_{p}%
\text{,}  \label{v1}
\end{equation}%
where the quantity $v_{p}$ has the meaning of the peculiar velocity with
respect to the particle flow \cite{zero}.

For the case of $e_{0}=1$ we return to $V^{(1)}=v_{st}+dr/d\tilde{t}$. Here, 
$v_{st}$ is the velocity in the static frame $v_{st}=\frac{P_{0}}{e_{0}}$%
(see, e.g. e. 43 in \cite{fl}). This means that for $e_{0}=1$ the coordinate
velocity $dr/d\tilde{t}$ is naturally decomposed into the flow velocity $%
v_{st}$ and the physical velocity $V^{(1)}$ as $dr/d\tilde{t}=V^{(1)}-v_{0}$%
. This decomposition resembles the Galilean summation law, but is valid
independently of how big the velocities are with respect to the speed of
light. Since the left hand side of this decomposition is a coordinate (not a
physical) velocity, it can be even superluminal. In the general case such a
decomposition is valid only for the rate of change of the proper distance $l$
with the time $\tilde{t}$ and for the flow velocity with respect to the
stationary frame $v_{st}$ since still $dl/d\tilde{t}=V^{(1)}-v_{st}$. (Here,
it is implied that the distance is measured along the hypersurface $\tilde{t}%
=const$ in the metric (\ref{met})$.$)

For a free particle it follows from (\ref{ut}), (\ref{ur}), (\ref{v1}) and (%
\ref{rtt}) that%
\begin{equation}
V^{(1)}=\frac{P_{0}e-Pe_{0}}{e_{0}e-PP_{0}}.  \label{v1e}
\end{equation}%
This is the general formula for radial motion, giving the 3-velocity of the
particle with the energy $e$ with respect to a free falling frame with the
energy $e_{0}$. Eqs. (\ref{Vg}), (\ref{v1e}) agree with eqs. (9), (31) of 
\cite{vm} and are more general in that they allow negative energy and
outward motion for particles and frames as well. Moreover, they can be
easily generalized to non-radial motion of a particle provided the frame is
still a radial one (see the next section).

It follows from (\ref{v1e}) that%
\begin{equation}
P=\frac{e(P_{0}-e_{0}V^{(1)})}{e_{0}-P_{0}V^{(1)}}\text{.}  \label{pv}
\end{equation}%
Taking the square of (\ref{pv}) and writing $P^{2}=e^{2}-e_{0}^{2}+P_{0}^{2}$
where (\ref{P0}), (\ref{dr}) are taken into account, we obtain

\begin{equation}
e=\gamma (e_{0}-P_{0}V^{(1)})\text{, }\gamma =\frac{1}{\sqrt{1-\left(
V^{(1)}\right) ^{2}}}\text{.}  \label{ep}
\end{equation}

When $e_{0}=1$, $P_{0}=v=\sqrt{1-f}$, the known formulas are reproduced.
Namely, eqs. (\ref{v1e}), (\ref{pv}) turn into eqs. (55), (56) of \cite{fl}
and eq. (\ref{ep}) turns into eq. (6.8) of \cite{zero}.

It is remarkable that despite the fact that the GP coordinates (\ref{met})
have no smooth limit for $e_{0}\rightarrow 0$, corresponding formulae for
3-velocities are regular for $e_{0}=0$.

\section{Motion with angular momentum and horizon asymptotics}

In the most general case of a free fall a particle can have nonzero angular
momentum $L$. Now, 
\begin{equation}
\frac{dr}{d\tau }=-P  \label{PL}
\end{equation}%
with%
\begin{equation}
P=\pm \sqrt{e^{2}-f(1+\frac{\mathcal{L}^{2}}{r^{2}})}\text{,}  \label{P}
\end{equation}%
$\mathcal{L=}\frac{L}{m}=u_{\phi }$, so 
\begin{equation}
\frac{d\phi }{d\tau }=\frac{\mathcal{L}}{r^{2}}.  \label{phi}
\end{equation}

As is known, in the central field a particle moves within a plane. If we
choose this plane to be $\theta =\frac{\pi }{2}$, 
\begin{equation}
h_{(3)\mu }=r(0,0,0,1)
\end{equation}%
and 
\begin{equation}
h_{(3)\mu }u^{\mu }=ru^{\phi }.
\end{equation}%
For the reference particle, eqs. (\ref{h0}), (\ref{hoc}) are valid. If we
take also into account (\ref{til}), this gives us 
\begin{equation}
\frac{d\phi }{d\tilde{t}}=\frac{\mathcal{L}f}{r^{2}(e_{0}e-PP_{0})},
\end{equation}%
\begin{equation}
V^{(3)}=\frac{\mathcal{L}f}{r(e_{0}e-PP_{0})}.
\end{equation}%
As for the radial component, it follows from (\ref{PL}), (\ref{P})\ that we
still have

\begin{equation}
V^{(1)}=\frac{P_{0}e-Pe_{0}}{e_{0}e-P_{0}P}  \label{V1}
\end{equation}%
If $e_{0}=1$, $P_{0}=\sqrt{1-f}\equiv v$, and we return to eq. (55) of \cite%
{fl}.

If $e_{0}=0$, 
\begin{equation}
V^{(1)}=-\frac{e}{P}\text{,}  \label{V10}
\end{equation}%
\begin{equation}
V^{(3)}=\frac{\mathcal{L}\sqrt{-f}}{rP}.
\end{equation}

The general formulae for the 3-velocity have several important limits. Now,
we consider the behavior at the horizon.  First we assume that $P$ and $P_{0}
$ have the same signs (say, positive, for definiteness). Let $f\rightarrow 0$%
, then 
\begin{equation}
P=\left\vert e\right\vert -\frac{f}{2\left\vert e\right\vert }(1+\frac{%
\mathcal{L}^{2}}{r^{2}})+...
\end{equation}%
\begin{equation}
P_{0}=\left\vert e_{0}\right\vert -\frac{f}{2\left\vert e_{0}\right\vert }%
+...
\end{equation}%
Substituting into the expressions for velocity components we can identify
two different cases. If $e$ and $e_{0}$ have the same sign, then

\begin{equation}
V^{(1)}\rightarrow \frac{e_{0}^{2}(1+\frac{\mathcal{L}^{2}}{r_{g}^{2}})-e^{2}%
}{e_{0}^{2}(1+\frac{\mathcal{L}^{2}}{r_{g}^{2}})+e^{2}}  \label{V1H}
\end{equation}%
and%
\begin{equation}
V^{(3)}\rightarrow \frac{2ee_{0}\mathcal{L}}{r_{g}[e^{2}+e_{0}^{2}(1+\frac{%
\mathcal{L}^{2}}{r_{g}^{2}})]}.  \label{V3H}
\end{equation}

If $e$ and $e_{0}$ have different signs, then independently of particular
values of the both energies, 
\begin{equation}
|V^{(1)}|\rightarrow 1,  \label{v=c}
\end{equation}%
\begin{equation}
V^{(3)}\rightarrow 0.  \label{v=0}
\end{equation}%
The same limits exist also if any of energies equals zero.

We can also see that if signs of $e_{0}$ and $e$ coincide, but the direction
of motion is different ($P_{0}$ and $P$ have different signs), the
asymptotic (\ref{v=c}) and (\ref{v=0}) holds as well.

All these cases are summarized in the Table II.

\bigskip 
\begin{tabular}{|l|l|l|l|l|}
\hline
Frame / Particle & $e>0,$ $P>0$ & $e<0,$ $P>0$ & $e>0$, $P<0$ & $e<0$, $P<0$
\\ \hline
$e_{0}>0$, $P_{0}>0$ & $V_{H}^{(1)}$; $V_{H}^{(3)}$ & $+1$; $0$ & $+1$; $0$
& -- \\ \hline
$e_{0}<0,$ $P_{0}>0$ & $-1$; $0$ & $-V_{H}^{(1)}$; $V_{H}^{(3)}$ & -- & $-1$%
; $0$ \\ \hline
$e_{0}>0$, $P_{0}<0$ & $-1$; $0$ & -- & $-V_{H}^{(1)}$; $V_{H}^{(3)}$ & $-1$%
; $0$ \\ \hline
$e_{0}<0$, $P_{0}<0$ & -- & $+1$; $0$ & $+1$; $0$ & $V_{H}^{(1)}$; $%
V_{H}^{(3)}$ \\ \hline
\end{tabular}

Here $V_{H}^{(1)}$ and $V_{H}^{(3)}$ are given by (\ref{V1H}) and (\ref{V3H}%
) respectively. For the cases where the frame and the particle can exist
only in non-intersecting zones of the Carter-Penrose diagram, the velocity
can not be defined.

As for critical particles or frames, the angular velocity at a horizon
always vanishes and the absolute value of radial velocity tends to $1$. As
for the sign of the radial component, $sign(V_{H}^{(1)})=sign(e_0/P_0)$ for
a critical particle ($e=0$, $e_0 \ne 0$) and $sign(V_{H}^{(1)})=-sign(e/P)$
for a critical frame ($e_0=0$, $e \ne 0$).

\section{The case V=1 and behavior of the Lemaitre time}

In a black hole space-time the most natural free-falling frames are
contracting ones realized by particles with a positive energy. A particular
case of $e_{0}=1$ (the Lema\^{\i}tre frame) is the most convenient choice
for such a frame, since constant time slices in this case are flat that
leads to some simplification. Regularity of the Lema\^{\i}tre frame at the
horizon makes the velocity with respect to this frame a meaningful
characteristic of particle motion, including the moment of horizon crossing.

To stress this point, let us consider for a moment the velocity with respect
to a \textit{stationary} frame at a horizon crossing. It is known that this
velocity is equal to the speed of light. However, the stationary frame at a
horizon becomes singular, so the equality $v_{st}=1$ tells us nothing about
a concrete properties of particle motion. On the contrary, the velocity with
respect to a frame which can be realized by massive particles contains
information about motion itself. We have seen that an ingoing positive
energy particle can have at a horizon any velocity from $0$ to $1$ depending
on the specific energy of the particle.

Meanwhile, we identified several cases when the velocity at the horizon
crossing with respect to \textit{Lema\^{\i}tre} frame is equal to the speed
of light. In the Schwarzschild space-time this happens for an outgoing
particle in the $R$ region and for negative energy particle in the $T$
region of the black hole. Moreover, the property of the velocity to be equal
to the speed of light depends only on signs of terms entering (\ref{v1e}),
so this property holds for any $e_0$-Lemaitre frame provided $e_0>0$. This
deserves some attention.

One can look at these features from somewhat different point of view. The
matter under discussion can be considered in the context of the Ba\~{n}%
ados-Silk-West (BSW) effect \cite{ban}. It states that the energy in the
center of mass frame $E_{c.m.}$ of two particles colliding near the black
hole horizon, under certain conditions can grow unbounded. Kinematically,
this means that the relative velocity of two particles tends to the speed of
light \cite{k}. Now, the role of one of particles is played by a reference
one. For a static particle, $E_{c.m.}$ does grow unbounded but for the frame
itself becomes singular. For free falling particles, the case when $e_{0}>0$
and $e>0$ corresponds to particle collisions of usual (without fine-tuning)
particles and cannot lead to the BSW effect \cite{ban}. This can be also
seen from the results of \cite{vm} where it was shown that the relative
velocity of particles moving from the $R$ region towards the nonextremal
horizon remains separated from the speed of light. Meanwhile, if $e_{0}>0$
but $e<0$, this corresponds to head-on collision that does lead to the
indefinite growth of $E_{c.m.}$ \cite{ps}, \cite{cqg}. However, this is
connected with a special role of the bifurcation point which will be
discussed elsewhere.

We discussed the case when the latter scenario is realized inside the
horizon in region II near the right branch of the future horizon. The
completely similar situation is possible near the left branch. Also,
collisions of such a type can occur in region IV where they have meaning of
collisions in the background of white holes \cite{white grib}, \cite{white}.

We could expect that something should prevent a realization of $V=1$.
Otherwise, the energy $E_{c.m.}$ would become not only unbounded but
infinite in the literal sense that is impossible \cite{cens}. A possible
reason appears to be clear when we consider the Lema\^{\i}tre time needed to
reach the horizon. Using (\ref{rtt}) it can be written as 
\begin{equation}
\Delta \tilde{t}=\int_{r_{0}}^{r_{h}}\frac{dr(e_{0}e-P_{0}P)}{Pf}.
\label{Lt}
\end{equation}

Let us calculate the $e_{0}$-Lemaitre time needed to reach a horizon. In
general, the right hand side of (\ref{Lt}) diverges at a horizon. However,
if both $e_{0}$ and $e$ are positive, as well as $P_{0}$ and $P$, the
divergent part in the integral cancels out and the time is finite. From
another side, changing sign of entities entering (\ref{Lt}) leads to
diverging result. For example, changing the sign of $P$ leads to%
\begin{equation}
\frac{dr}{d\tau }=+v=\sqrt{e^{2}-f}
\end{equation}%
and 
\begin{equation}
\frac{d\tilde{t}}{d\tau }=\frac{e_{0}e+P_{0}\left\vert P\right\vert }{f}
\label{dt2}
\end{equation}%
instead of (\ref{dt}), so that we have summations of divergent parts in (\ref%
{Lt}).

We can identify three more situations with infinite time $\tilde{t}$.
Indeed, instead of reversing the sign of $P$ the same effect can be got by
reversing the sign of $e$ , the sign of $e_{0}$ or the sign of $P_{0}$.

So that, by inspecting the Lema\^{\i}tre time in mentioned above the two
situations, we have seen that the horizon crossing leading to $V=1$ happens
for infinite Lema\^{\i}tre time (for the above cases, actually, in
infinitely remote past). However, the reference particle or an observer
comoving with the frame with respect to which the velocity is measured,
reaches the horizon for finite Lema\^{\i}tre time. Therefore, they cannot
meet in the same point of space-time, measurement of the velocity (as well
as collision) is impossible, so the value $V=1$ and infinite $E_{c.m.}$
remain unreachable. Moreover, if the velocity of a particle reaches $1$ with respect to
any other reference system from the Table II, the time of this system needed for the particle 
to reach horizon diverges, and the above argument is valid as well.

It is worth noting that the velocity of a particle approaches the speed of
light just on the boundary where a corresponding frame covers a part of
space-time, so its incompleteness reveals itself. For the stationary
observer, this is a horizon beyond which a static frame cannot be extended
and where the acceleration tends to infinity. For a free-falling frame, the
horizon separates different region where motion with a given sign of $e$ or $%
P$ is possible or not.

\subsection{Special case: critical particles and frames\label{cr}}

In this subsection we present notes about critical frames ($e_{0}=0$) and
particles ($e=0$) which are curious by themselves. Let us consider the time
needed to fall into a singularity from a horizon. It is equal to 
\begin{equation}
\Delta \tilde{t}=\int_{0}^{r_{h}}\frac{dr(e_{0}e-P_{0}P)}{Pf}%
=\int_{0}^{r_{h}}\frac{dre_{0}e}{Pf}-\int_{0}^{r_{h}}\frac{dr\sqrt{%
e_{0}^{2}-f}}{f}.  \label{time}
\end{equation}

We can see that 1) If $e_{0}=0$,%
\begin{equation}
\Delta \tilde{t}=\int_{0}^{r_{h}}\frac{dr}{\sqrt{-f}}
\end{equation}%
does not depend on $e$ and, correspondingly, on details of the the
observer's motion. This property is evident in the coordinate system
obtained from static coordinates via signature change (\ref{g}), when the
spatial and temporal coordinates interchange. Indeed, the metric in such a
system, being a particular case of the Kantowski-Sachs cosmological metric
depends on time but not on a spatial coordinate, so that a singularity is in
future for any observer, separated from the observer by some time which does
not depend on the actual position of the observer. What is remarkable, is
that the limit $e_{0}\rightarrow 0$ is smooth since the first term in (\ref%
{time}), which does depend on details of the observer trajectory, vanishes
smoothly for $e_{0}\rightarrow 0$.

2) If $e_{0}\neq 0$ but $e=0$ \ (the critical particle)%
\begin{equation}
\Delta \tilde{t}=\int_{0}^{r_{h}}\frac{dr\sqrt{e_{0}^{2}+g}}{g}.
\end{equation}

Near $g=0$ it diverges as $\Delta \tilde{t}\sim \left\vert \ln
(r_{h}-r)\right\vert .$ This means that for any critical particle the
corresponding time $\tilde{t}$ of fall is infinite in all frames with $%
e_{0}\neq 0$.

\section{Singularity or turning points}

Going further, we consider asymptotics near the singularity. For the
Schwarzschild metric, it is reachable. We consider the black hole region,
assuming that the frame is contracting and a particle is ingoing, so $P>0$, $%
P_{0}>0$. Then, $f\approx -\frac{2M}{r}\rightarrow -\infty $. $P\approx 
\frac{\sqrt{-f}\left\vert \mathcal{L}\right\vert }{r}$, $P_{0}\approx \sqrt{%
-f}.$

If $L\neq 0$,%
\begin{equation}
V^{(3)}\rightarrow sign\mathcal{L}\text{,}
\end{equation}%
\begin{equation}
V^{(1)}\approx \frac{e_{0}}{P_{0}}\approx \frac{e_{0}}{\sqrt{-f}}\rightarrow
0
\end{equation}%
for any finite $e_{0}\neq 0$ and $e$.

If $e_{0}=0$%
\begin{equation}
V^{(1)}=-\frac{e}{P}\approx -e\frac{r}{\sqrt{-f}\left\vert \mathcal{L}%
\right\vert }\rightarrow 0\text{.}
\end{equation}

If $L=0$, $V^{(3)}=0$, $P\approx \sqrt{-f}\approx P_{0}$%
\begin{equation}
V^{(1)}\approx \frac{e_{0}-e}{\sqrt{-f}}\rightarrow 0
\end{equation}
The formulas agree with \cite{fl}, Sec. 7 if we put $e_{0}=1$.

For certain space-times a singularity is unreachable. The most common
example is the Reissner-Nordstr{\"o}m (RN) black hole with $%
f=1-r_{g}/r+Q/r^{2}$. (As detailed consideration of the RN metric is beyond
the scope of the present paper, we restrict ourselves by several short
notes.) For the electric charge $Q$ small enough, this metric contains two
horizons, with the $T$ region located between them. In the RN black hole the
singularity is unreachable. The results of Sec. VI tells us that the $e_{0}$%
-Lema\^{\i}tre time ($e_{0}>0$) needed for a particle with $e\leq 0$ to
cross the entire T-region diverges. Only particles with $e>0$ can cross the
inner horizon. Then, they have a turning point at 
\begin{equation}
P=0\text{, }e^{2}=f
\end{equation}%
with 
\begin{equation}
V^{(1)}=\frac{P_{0}}{e_{0}}=\frac{\sqrt{e_{0}^{2}-e^{2}}}{e_{0}}
\end{equation}%
and 
\begin{equation}
V^{(3)}=\frac{\mathcal{L}f}{re_{0}e}=\frac{e\mathcal{L}}{e_{0}r}.
\end{equation}

Only frames with $e_{0}>e$ can describe the motion of a test particle with
energy $e$ till the turning point. After the turning point, the motion turns
to be outward, towards the inner horizon from the $R$ region inside. We see
from results of Sec VI that for a particle after the turning point the $%
e_{0} $-Lema\^{\i}tre time needed to reach the inner horizon again is
infinite for any contracting free falling reference system.

\section{Discussion and Conclusions}

In the present paper we have considered the formulae for the velocity with
respect to the most general free falling radially frame in a spherically
symmetric space-time. The formulae can be applied to any radially falling
reference system including expanding systems, and those realized by
particles with negative energy. The particle moving in corresponding
backgrounds can be, in its turn, ingoing or outgoing, having positive or
negative energy.

The velocity properties with respect to a generalized frame realized by
ingoing particles with a positive energy $e_{0}\neq 1$ differs from those
for the original frame ($e_{0}=1$) only quantitatively. If there are no
special reasons, the $e_{0}=1$ frame is preferable since it gives a number
of simplifications. However, we can indicate certain cases when a $e_{0}\neq
1$ frame can be useful and can be not only of pure academic interest. All of
them deal with non-Schwarzschild space-times, and there are the cases when
the standard $e_{0}=1$ reference system fails to describe a motion \cite%
{Faraoni}. For example, in the RN black hole the $e_{0}=1$ frame can not be
prolonged through the turning point at $f=1$. So that a motion of a particle
with $e>1$, having its turning point at $f=e^{2}$ cannot be fully covered by
the Lema\^{\i}tre frame with $e_{0}=1$, but can be covered in any frame with 
$e_{0}>e$. Another example deals with the Schwarzschild - de Sitter metric
with $f=1-r_{g}/r-\Lambda r^{2}$, where $\Lambda >0$ is the cosmological
constant. Now the function $f(r)$ has a maximum value $f_{max}<1$, and the
flow velocity of a (would be) Lema\^{\i}tre frame $v=\sqrt{1-f}$ never
crosses zero. This means that such a frame is everywhere contracting.
 By
reversing sign of $v$ we can get also an everywhere expanding frame, both
are  unsuitable to describe a black hole in expanding de Sitter Universe.
However, taking $e_{0}^{2}=f_{max}$ we can construct a smooth  frame expanding for
large $r$ and contracting for small $r$ with the stationary point at the maximum of $f(r)$
where contracting and expanding branches are glued together (for close but somewhat different approach to this
particular metric see the recent paper \cite{gr}).

It is not so easy to imagine a situation where expanding frames, or frames
realized by negative energy particles could be preferable (apart from such
an exotic as a white hole, for which an expanding frame is as natural as
contracting frame for a black hole!), nevertheless, the ability of our
general formulae to include these frame as well is rather interesting.

We have shown also that non-zero angular momentum of a particle can be
included in the general scheme rather easily if the frame is still radially
moving.

\section*{Acknowledgements}

The work of AT is supported by the Program of Competitive Growth of Kazan
Federal University and by the Interdisciplinary Scientific and Educational
School of Moscow University in Fundamental and Applied Space Research.

\end{document}